\documentclass[12pt]{article}
\usepackage{latexsym,epsfig,amssymb,euscript,amsmath,verbatim}

 \def\be{\begin{equation}} \def\ee{\end{equation}}
\def\bea{\begin{eqnarray}} \def\eea{\end{eqnarray}}
\def\ie{{\em i.e.}}
\newcommand{\nn}{\nonumber}

\begin{document}

\begin{titlepage}
\begin{flushright}
SISSA 18/2010/EP\\
IFUP-TH/2010-12

\end{flushright}
\bigskip
\def\thefootnote{\fnsymbol{footnote}}

\begin{center}
{\Large {\bf
On Holographic Superconductors \\ \vskip 12pt with DC Current
  } }
\end{center}

\bigskip
\begin{center}
{\large Daniel Are\'an$^1$, Matteo Bertolini$^1$,
Jarah Evslin$^2$ \\ \vskip 5pt and Tom\'a \hskip -4pt\v{s} Proch\'azka$^1$}
\end{center}

\renewcommand{\thefootnote}{\arabic{footnote}}

\begin{center}
\vspace{1em}
${}^1\;$SISSA and INFN - Sezione di Trieste \\
Via Bonomea 265; I-34136 Trieste, Italy \\
\vspace{1em} ${}^2\;$Department of Physics, University of Pisa and INFN - Sezione di Pisa\\ Largo
Pontecorvo 3, Ed. C; I-56127 Pisa, Italy

\vspace{1em}
\texttt{arean@sissa.it, bertmat@sissa.it, jarah@df.unipi.it, procht@sissa.it}

\end{center}

\noindent
\begin{center} {\bf Abstract}
\end{center}
We study direct currents in a simple holographic realization of a
superconducting film. We investigate how the presence of a DC current
affects the superconducting phase transition, which becomes
first order for any non-vanishing value of the current, as well as several
other properties of the superconductor such as the AC conductivity.
Near the critical temperature we find a quantitative agreement with several
properties of Ginzburg-Landau superconducting films, for example the squared ratio
of the maximal and minimal condensate is equal to two thirds.
We also comment on the extension of our construction to holographic Josephson
junctions.

\vfill

\end{titlepage}

\hfill{}

%\tableofcontents

\setcounter{footnote}{0}

%%%%%%%%%%%%%%%%%%%%%%%%%%%%%%%%%%%%%%%%%%%%%%%%%%%%%%%%%%%
%%%%%%%%%%%%%%%%%%%%%%%%%%%%%%%%%%%%%%%%%%%%%%%%%%%%%%%%%%
\section{Introduction}

In \cite{Gubser:2008px} it was shown that a black hole solution in a theory with a charged scalar field coupled
to Maxwell-Einstein gravity
may become classically unstable below some critical temperature $T_c$. This instability
induces charged scalar hair for the black hole for $T<T_c$. According
to the AdS/CFT duality, the holographic dual of such a system is a thermal quantum field theory in flat Minkowski
space with a global $U(1)$ symmetry, which is spontaneously broken below $T_c$ by the condensation of the operator
dual to the bulk complex scalar. In this sense, the boundary theory has many of the necessary ingredients to describe a
superconductor, or a superfluid \cite{weinberg}. This result was first exploited in \cite{Hartnoll:2008vx} to assemble
a gravity dual of a system undergoing a superconducting phase transition. This construction has been widely studied and
generalized, and the different gravity duals sharing these same basic features go under the name of holographic
superconductors (see \cite{Hartnoll:2009sz,Herzog:2009xv,Horowitz:2010gk} for reviews and references).

In a holographic superconductor, a background magnetic field induces a
current. However, because of the absence of a dynamical gauge
field,  this current does not expel the magnetic field, unlike the
current induced in ordinary superconductors.  In this sense
holographic superconductors more closely resemble thin superconducting
films or wires.  Motivated by this analogy, in the present note we
will compare and contrast holographic superconductor phenomenology
with that of superconducting films. More precisely, we study a holographic superconductor in two
spatial dimensions, {\it i.e.} a (extremely) thin superconducting film,
with a DC current, analyzing its phase diagram and the behavior of some
interesting thermodynamic quantities.

In fact, such a system is interesting to study for a second, more ambitious,
reason. It is believed that holographic superconductors may give an understanding of some basic
features of high temperature superconductors (HTS). HTS typically enjoy a layered structure
and, according to the Lawrence-Doniach model \cite{lawrence}, may be approximated by films of superconductors
separated by Josephson junctions. Therefore, a holographic
realization of a Josephson junction would be desirable. The latter is based on the
Josephson effect \cite{josephson}, the phenomenon of current flow
across two weakly coupled superconductors separated by a very thin
insulating barrier (the Josephson junction). Our model can then be seen
as a first, necessary step towards the realization of a Josephson junction.

We pursue the phenomenological approach of \cite{Hartnoll:2008vx}
and therefore the gravitational system we consider is Einstein-Maxwell
theory in four dimensions minimally coupled to a charged massive
scalar field.  In \cite{Hartnoll:2008vx} the following basic set up was
considered: the condensation of the charged scalar in a black hole
metric at finite charge density. The black hole introduces a
temperature $T$. The finite charge density, which is taken care of by
allowing for a non trivial profile for the temporal component $A_0$ of
the gauge field, provides an independent scale needed to get a
critical temperature $T_c$. One then finds that for temperatures below
$T_c$ the charged scalar condenses.  Under the AdS/CFT map gauge
symmetries on the gravity side are dual to global ones on the field
theory. Then the condensation of the scalar nicely realizes the
spontaneous breaking of a global $U(1)$ symmetry.

We want to modify this basic scenario and
allow for the presence of a DC current. To this end, we consider
solutions where also a spatial component of the gauge field has a non
trivial profile, this providing, via the AdS/CFT map, a current in
the dual theory. Such solutions can be easily found in the
superconducting phase, where the scalar is non-zero. However, as
pointed out in \cite{Basu:2008st}, in the normal state (where the
symmetry is not broken and the scalar is hence vanishing) the only
allowed solutions for the spatial components of the gauge field are
the trivial ones. For this reason, within the minimal Einstein-Maxwell
framework, we cannot construct a model describing the normal state
with DC current. However, as we will discuss in detail, this
inconvenience will not impede us to obtain some robust results
characterizing the behavior of holographic superconductors at fixed DC
current.

An additional limitation of our approach comes from the fact that, as
in \cite{Hartnoll:2008vx}, we work in the probe approximation. This is
the limit where the backreaction of the gauge and scalar fields on
the metric is neglected. Hence our results are reliable only in the
regime where the backreaction can be effectively neglected.  Luckily,  
while the probe approximation breaks down in the zero temperature limit, for temperatures
significantly different from zero the results obtained in the probe
limit are not substantially modified by the backreaction \cite{Hartnoll:2008kx}. There is
therefore a large region where our results should not be sensibly
different from a honest fully back-reacted model\footnote{In
\cite{Gauntlett:2009dn,Gubser:2009qm}, it has been shown that a
phenomenological model of the likes of \cite{Hartnoll:2008vx} can be
consistently embedded in M-theory or type IIB String Theory. Such
embeddings constitute an important advance towards the understanding
of the underlying microscopic theory of the holographic
superconductors. Unfortunately, sticking to the probe approximation
prevents us from using the models of
\cite{Gauntlett:2009dn,Gubser:2009qm} since the charge there is fixed
to a finite value, while the probe approximation holds in the large
charge limit.}.

The same model we are going to study here was already considered in
\cite{Basu:2008st,Herzog:2008he}. Differently from those analyses we
study the system at fixed current. This choice, beside being closer in
spirit to real-life experiments, allows us to obtain new results about
the phase diagram of this system together with interesting checks and
predictions for the behavior of holographic superconductors with DC
current, as we now summarize:

\begin{itemize}

\item At any finite DC current the transition between the
superconducting and the normal state is a first order phase
transition. We study the temperature dependence of the condensate and
compute the free energy in the superconducting state, concluding that
at the phase transition the condensate always jumps a finite distance
to zero. This is a clear indication of a first order phase transition.

\item We determine the relation between the current and the superfluid
velocity. We largely find nice agreement with expectations for
physical superconducting films, both for temperatures appreciably
lower than, and close to, $T_c$. Moreover, it turns out that the form
of these curves further justifies the assertion in the previous point,
namely that the phase transition is first order. Interestingly,
at low temperatures we find that, in contrast with BCS superconducting
films, for each value of the superfluid velocity there are two possible
values of the current. A free energy computation then shows that only
one value, in fact the highest one, is thermodynamically stable.

\item We study the temperature dependence of the critical current and
of the ratio given by the value of the condensate at zero current over
the value at the critical current. For temperatures close to $T_c$ we
find that the holographic superconductors reproduce the universal
results predicted by the Ginzburg-Landau (GL) model for
superconducting films.  On the other hand, at lower temperatures our
results deviate significantly from the ones of GL.

\item Finally, we study the dependence of the AC conductivity on the
DC current. We present results for the conductivity in the direction
transverse to the current. At low temperatures we analyze the
dependence of the frequency gap on the DC current. As expected on
physical grounds, as we increase the current the frequency gap
diminishes, and it does so down to a minimal (but not vanishing) value
where the first order phase transition occurs.

\end{itemize}

The rest of this note is organized as follows. In section 2 we present
the bulk Lagrangian, the equations of motion which we have solved numerically, and
motivate our ansatz and boundary conditions. In section 3 we present
the physical output of our numerical studies, namely the checks and
predictions mentioned above. Section 4 contains our conclusions
as well as a possible strategy for constructing a holographic dual of (an array
of) Josephson junctions.

%%%%%%%%%%%%%%%%%%%%%%%%%%%%%%%%%%%%%%%%%%%%%%%%%%%%%%%%%%%%%%%%%%%%%%%%%%%%%%%%%%%%%%%%
%%%%%%%%%%%%%%%%%%%%%%%%%%%%%%%%%%%%%%%%%%%%%%%%%%%%%%%%%%%%%%%%%%%%%%%%%%%%%%%%%%%%%%%%
\section{The gravity dual of a DC superconductor }

As advertised, we pursue a bottom-up approach to holographic
superconductivity, and consider as a starting point the model
originally presented in \cite{Hartnoll:2008vx}, Einstein-Maxwell
theory in 4-dimensions minimally coupled to a charged, massive scalar
field. We stick to the probe approximation and in this case the action
of the scalar-Einstein-Maxwell theory reduces to
\be
\mathcal{S}= \int dx^{\,4}\,\sqrt{-g}\,
\left[-\frac{1}{4}F_{\mu\nu}F^{\mu\nu}-|(\partial_\mu-iA_\mu)\Psi|^2-m^2\,\Psi^* \Psi \right]~,
\label{action}
\ee
where the Einstein-Hilbert term has been suppressed, since the
backreaction of the fields on the metric can be ignored in the probe
limit (the Einstein equations decouple)\footnote{ Let us start from
the standard Lagrangian
$\sqrt{-g}\,\left[-\frac{1}{4}F_{\mu\nu}F^{\mu\nu}-|
(\partial_\mu-i\,q\,A_\mu)\Psi|^2-m^2\,|\Psi|^2
\right]\,+\,{\rm Einstein}$, rewrite it in terms of the rescaled
fields $\tilde\Psi=q\,\Psi$ and $\tilde A_\mu= q\,A_\mu$ and take the
limit $q\to\infty$ while keeping $\tilde \Psi\,,\;\tilde A_\mu$
fixed. Then, the Lagrangian becomes ${1\over
q^2}\,\sqrt{-g}\,\left[-\frac{1}{4}\tilde F_{\mu\nu}\tilde
F^{\mu\nu}-|(\partial_\mu-i\,\tilde A_\mu)\tilde
\Psi|^2-m^2\,|\tilde\Psi|^2 \right]\,+\,{\rm Einstein}$.  Due to the
$1/q^2$ factor the matter sources decouple from the Einstein equations
and the dynamics of the vector and the scalar field are described by
the action (\ref{action}) in a vacuum solution of the Einstein
equations.}.  $F_{\mu\nu}$ is the $U(1)$ field strength, $\Psi$ is the
complex scalar with charge 1 and mass $m$, and $g$ is the determinant
of the metric $g_{\mu\nu}$, which we take to be the asymptotically AdS
planar black hole metric \be
\label{metric}
ds^2=-f(r)dt^2+\frac{dr^2}{f(r)}+{r^2\over
L^2}(dx^2+dy^2)~~~~~\mbox{where}~~~ f(r)=\frac{r^2}{L^2}-\frac{M}{r}~.
\ee The radial direction extends from the black hole horizon at $r=
r_0=(ML^2)^{1/3}$ to the boundary of AdS at $r\rightarrow\infty$, $L$
is the radius of AdS and $M$ the mass of the black hole.  Beside the
holographic coordinate $r$, we have three others ($t,x,y$), which
parametrize the AdS boundary and hence the (2+1)-dimensional dual
field theory space-time.

The temperature of the black hole (and hence of the dual field theory)
is given by \be T={3\over4\pi\,L^2}\,r_0~.
\label{temp}
\ee As in \cite{Hartnoll:2008vx} we will take the scalar mass to be
$m^2=-2/L^2$ which is above the Breitenlohner-Friedman bound.

%%%%%%%%%%%%%%%%%%%%%%%%%%%%%%%%%%%%%%%%%%%%%%%%%%%%%%%%%%%%%%%%%%%%%%
\subsection{The ansatz}

According to the AdS/CFT map, the VEV of the $U(1)$ current in the
dual field theory is identified with the subleading boundary
asymptotics of the bulk gauge field. Hence, to describe
holographically a superconductor with a DC current, we need to look
for bulk solutions where the black hole develops charged scalar hair
in the presence of a non-trivial profile for a spatial component of
the gauge field.  More precisely, we are interested in getting a
current in the $x$ direction, therefore we will look for solutions
which are independent of the time coordinate $t$ and of $y$, but with
a non-trivial dependence on both $r$ and $x$.  We choose the gauge
$A_r=0$ (this leaves the freedom to perform $r$-independent 
gauge transformations, as we will do later). As our solutions are $y$-independent, we set
$A_y=0$. Thus we must determine $A_x,\ A_t,$ and $\Psi$ as functions
of $x$ and $r$.

We choose the modulus of the scalar to be independent of $x$ and
similarly for $A_x$ and
$A_t$. However, having a current then requires that the phase of
$\Psi$ be $x$-dependent. The simplest such ansatz reads
\be
\Psi(r,x)
= \psi(r) \, e^{i\theta x}~,
\ee
which automatically satisfies the
equations of motion for $A_r$ and for the phase of $\Psi$, the latter
imposing that $\theta$ is indeed a constant. Summarizing, $\psi$, $A_x$
and $A_t$ are functions of  $r$
while $\theta$ is a constant. Notice that in a superconductor the
spatial derivative of the phase of the condensate is the superfluid
velocity \cite{Tinkham,Herzog:2008he}. Therefore, in our case we are
describing a superconductor with a constant superfluid  velocity,
since by the AdS/CFT  map the latter is given by
$\theta$.

Notice that a non-zero $A_x$ contributes positively to the effective
scalar mass, hence one expects that a sufficiently large $A_x$ will
win against the negative contribution coming from the time-component
of the gauge potential, eventually destroying black hole
superconductivity \cite{Basu:2008st}. This corresponds to a critical
maximal current in the dual field theory, above which the system
enters the normal phase, which is indeed what is expected for physical
superconductors.

The equations of motion following from the action (\ref{action}) are
\bea
\label{eom1}
\partial_{r}\left( r^2\, \partial_{r} A_t\right)
-2\,\frac{r^2\,\psi^2}{f}\,A_t&=&0~,\\
\label{eom2}
\partial_{r}\left(f\,\partial_{r}
A_x\right)-2\psi^2\,\left(A_x-\theta\right)&=&0~,\\
\label{eom3}
\partial_{r}\left(
r^2\,f\,\partial_{r}\psi\right)-L^2\left(A_x-\theta\right)^2\psi+\frac{r^2
A_t^2}{f}\, \psi+\frac{2 r^2}{L^2}\,\psi &=&0~.
\eea
These equations are invariant under two independent scaling symmetries
\bea
& r \rightarrow \lambda r ~,~ r_0 \rightarrow \lambda r_0 ~,~ L
\rightarrow L ~,~ \psi \rightarrow \psi ~,~ A_t \rightarrow \lambda
A_t ~,~ A_x \rightarrow \lambda A_x ~,~ \theta \rightarrow \lambda
\theta ~, \\
& r \rightarrow r ~,~ r_0 \rightarrow r_0 ~,~ L
\rightarrow \nu L ~,~ \psi \rightarrow \nu^{-1} \psi ~,~ A_t
\rightarrow \nu^{-2} A_t ~,~ A_x \rightarrow \nu^{-2} A_x ~,~ \theta
\rightarrow \nu^{-2} \theta~. \nn
\eea
When performing numeric computations we find it convenient to work with variables and
coordinates which are invariant with respect to these rescalings (and
hence dimensionless). In our case they are
\be
\frac{r}{r_0}~~,~~{L^2\over r_0}\,A_{\mu}~~,~~{L^2\over
r_0}\,\theta~~,~~L\,\psi~.
\label{newfields}
\ee
The equations of motion written in terms of these rescaled and
dimensionless quantities are the same as before  with $r_0$ and $L$
set equal to one.

%%%%%%%%%%%%%%%%%%%%%%%%%%%%%%%%%%%%%%%%%%%%%%%%%%%%%%%%%%%%%%%%%%%%%%%%%%%%%%%%%
\subsection{Asymptotics and their dual interpretation}

The equations of motion (\ref{eom1})-(\ref{eom3}) are second order,
and so we expect six constants of integration, which together with
$\theta$ imply seven parameters. Regularity at the horizon sets
$A_t=0$ at $r=1$ and this, via the equations of motion, imposes
two more constraints on $\psi$ and $A_x$. This leaves four parameters.
In addition, there are boundary conditions at the boundary of AdS. The
leading asymptotics of the fields at large $r$ read
\be
A_x=A_x^{(0)}-\frac{A_x^{(1)}}{r}+O(r^{-2})~~~,~~~
A_t=A_t^{(0)}-\frac{A_t^{(1)}}{r}+O(r^{-2})~~~,~~~
\psi=\frac{\psi^{(1)}}{r}+\frac{\psi^{(2)}}{r^2}+O(r^{-3})~,
\label{exp}
\ee
while $\theta$ is constant everywhere. The leading contribution of
the time and space components of the gauge field correspond, via the
AdS/CFT map, to a chemical potential $\mu$ and a source for the
$x$-component of the dual current, respectively. This source for the
current will be nothing else than the superfluid velocity $\nu_x$.
Since the gauge field is covariantly coupled to the scalar, on the
boundary we get a term of the form
$\partial_x\,\varphi-A_x^{(0)}=\theta-A_x^{(0)}$ and we see that one
can, through a gauge transformation, trade $\theta$ for $A_x^{(0)}$
\cite{Tinkham,Herzog:2008he}. In fact, from now on we will choose to
work in the gauge $\theta=0$ and identify the superfluid velocity with
$A_x^{(0)}$. The subleading asymptotics correspond instead to the
charge density $\rho$ and the VEV of the current density $J_x$.

With our choice of scalar mass term we have
two options for the corresponding dual operator. This is because both
asymptotic behaviors of the scalar are normalizable at the boundary,
so both of them can correspond to a VEV of a dual operator
\cite{Klebanov:1999tb}.  We can choose the leading asymptotic
coefficient $\psi^{(1)}$ to be a source of a scaling dimension $2$
operator $O_2$. In this case the expectation value of $O_2$ will be
proportional to the subleading coefficient $\psi^{(2)}$. Or, we can
choose $\psi^{(2)}$ to be a source of an operator $O_1$ with scaling
dimension $1$. In this case the VEV of $O_1$ will be proportional to
$\psi^{(1)}$. For definiteness, in what follows we will work with the
operator $O_2$ (the basic results are not qualitatively different for
the opposite choice). To have spontaneous breaking of the $U(1)$
symmetry, we want the source to vanish and hence we will impose
$\psi^{(1)}=0$.

In summary, we have seven parameters, three regularity conditions at
the horizon plus the two conditions $\psi^{(1)}=0~,~\theta=0$. Hence,
we expect a two-parameter family of solutions, which may be
parametrized, for example, by the temperature and current of the
superconducting film. Given the temperature and current one may then
calculate the value of the order parameter $\langle O_2\rangle$.

Undoing the rescaling described above, we can rewrite the field theory
quantities in terms of the asymptotic coefficients of the dimensionless
fields. We find \bea &&\mu={4\pi\over3}\,T\,A_t^{(0)}\,,\qquad
\rho=\langle J_t\rangle={16\pi^2\over9}\,T^2\,A_t^{(1)}\,,\nonumber\\
\nonumber\\
&&\nu_x={4\pi\over3}\,T\,A_x^{(0)}\,,\qquad
j_x\equiv\langle J_x\rangle={16\pi^2\over9}\,T^2\, A_x^{(1)}\,,\qquad
{\langle O_{2}\rangle}= 2 {16\pi^2\over9}\,T^2\,\psi^{(2)}~,\nonumber\\
\label{physquant}
\eea where we have used the field and coordinate redefinitions given
by eq.~(\ref{newfields}), and written $r_0$ in terms of the
temperature via eq.~(\ref{temp}).

As can be seen from the expressions for the chemical potential and
charge density in eq. (\ref{physquant}), the  asymptotic behavior of
$A_t$ only determines the dimensionless ratios $\mu/T$ and
$\rho/T^2$. In other words, the  gravity dual only gives us
information about the dimensionless ratio of the two scales in the
theory: the temperature  and the charge density (or the chemical
potential). One can then decide to use either the chemical potential
or the charge  density to fix a scale. We will use the former and
study the evolution of $\langle O_{2}\rangle/\mu^2$ as a function  of
$T/\mu$ and $j_x/\mu^2$. Accordingly, using eq.~(\ref{physquant}), we
define $T_c$, the critical temperature at zero current, as 
\be 
{T_c\over \mu}={3\over 4\pi}\,\frac{1}{A_t^{(0)}|_c}~, 
\ee
where $A_t^{(0)}|_c$ is the critical value of $A_t^{(0)}$ for which
the condensate turns on at $j_x=0$.  When studying the thermodynamics of the
system we will work in the grand canonical ensemble, which corresponds
to a system at fixed chemical potential.

%%%%%%%%%%%%%%%%%%%%%%%%%%%%%%%%%%%%%%%%%%%%%%%%%%%%%%%%%%%%%%%%%%%%%%%%%%%%%%%%%%%%%%%%
%%%%%%%%%%%%%%%%%%%%%%%%%%%%%%%%%%%%%%%%%%%%%%%%%%%%%%%%%%%%%%%%%%%%%%%%%%%%%%%%%%%%%%%%
\section{Holographic predictions}
\label{resultsect}

We have numerically solved the system of coupled differential
equations (\ref{eom1})-(\ref{eom3}) and determined the condensate
$\langle O_2\rangle$ as a function of the current and the
temperature. Using a shooting technique we have integrated the
equations from the horizon up to the boundary, with the boundary
conditions discussed before.  Via the AdS/CFT maps detailed in
eqs.~(\ref{physquant}), we have then determined the surface of
solutions for the condensate as a function of the current and the
temperature.

%%%%%%%%%%%%%%%%%%%%%%%%%%%%%%%%%%%%%%%%%%%%%%%%%%%%%%%%%%%%%%%%%%%%%%%%%%%%%%%%%%%%%%
\subsection{Phase transition with DC current}
\label{phasetranssect}

\begin{figure}
\begin{center}
\includegraphics[width=1\textwidth]{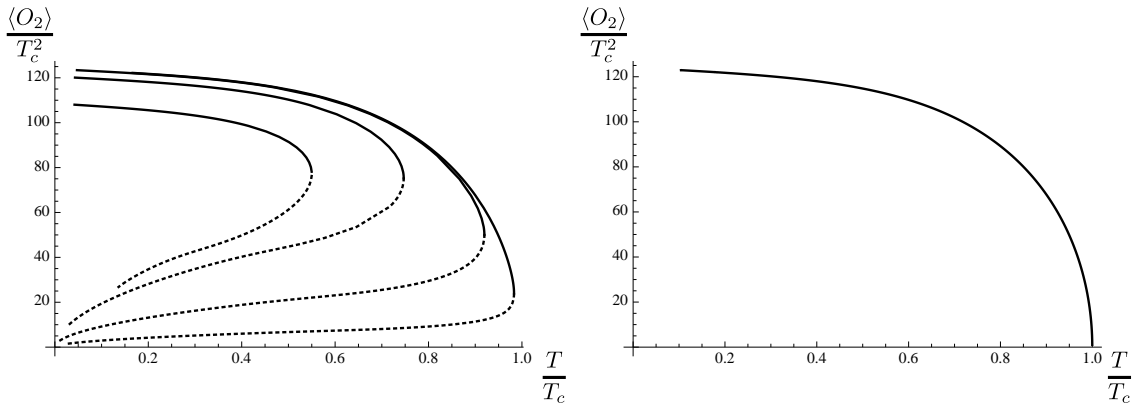}
\end{center}
\caption[Unflavor]{\small On the left we plot $\langle O_2\rangle$
versus the temperature for several values of the current: from the
innermost to the outermost
$j_x/T_c^2=28.98\,,\;14.49\,,\;2.90\,,\;0.290$. The dotted lines
correspond to the states with larger free energy than their
counterparts at the same temperature.  On the right we show for
comparison the result at zero current. Notice that at the critical
temperature $\langle O_2\rangle$ vanishes in this case.  }
\label{condvstemp}
\end{figure}

The first important thing we want to analyze is how the presence of
the current modifies the temperature dependence of the
condensate. This is shown in figure \ref{condvstemp} for different 
values of the current and compared with the result at zero
current obtained in \cite{Hartnoll:2008vx}. Two significant
modifications occur. First, at any finite value of the current one
can see that the curve $\langle O_2\rangle$ vs $T$ becomes bivaluated.
There appears a new branch (the dotted line) corresponding to states
where the value of the condensate is much lower. In the following, by computing the free
energy we will see that the states with lower value of the condensate
have a larger free energy than their counterparts with larger $\langle
O_2\rangle$ at the same temperature. Therefore, this new branch
corresponds to thermodynamically disfavored states
\footnote{Notice that this branch smoothly joins the (unstable)
normal phase branch of \cite{Hartnoll:2008vx} in the $J_x \rightarrow
0$ limit.}.  Second and more importantly, we observe that the
superconducting state exists up to a maximum value of the temperature
(where the plot turns back). Crucially, at that point the value of the
condensate is larger than zero. Therefore, at the phase transition the
condensate must jump a finite distance to zero. Unless one fine tunes the
parameters, such a jump will almost certainly change the energy and so
require some latent heat, implying that the phase transition is first order. Moreover, as 
is expected on physical grounds, the temperature at which the phase transition occurs is always 
lower than $T_c$, the critical temperature at zero current, and its value decreases with  
increasing current.

This phase transition pattern is quite different from that of
refs.~\cite{Basu:2008st,Herzog:2008he}. The analysis performed there
corresponds to experiments where instead of the current, the
superfluid velocity is kept fixed. There it was found that the
superconducting phase is separated from the normal phase by a second
order phase transition from zero superfluid velocity up to a
tricritical point where the phase transition becomes first order and
remains so up to the maximum velocity, where the phase transition
would be at zero temperature (similar results were found in
\cite{Basu:2008bh}, in the context of superconducting D-brane
models). In fact, as we will see in section \ref{jvsvelsec}, the
different phase transition pattern one finds when working at finite
current agrees with what is known about the relation between the
current $j_x$ and the superfluid velocity $\nu_x$ in superconducting
films.

%%%%%%%%%%%%%%%%%%%%%%%%%%%%%%%%%%%%%%%%%%%%%%%%%%%%%%%%%%%%%%%%%%%%%%%%%%%%%%%
\subsubsection{The free energy}
\label{freeensect}

In order to confirm our previous claim, namely that the states with
lower value of the condensate are metastable, we shall now
compute the free energy of the superconducting phase and show that it
is larger for the metastable branch (dotted line in figure
\ref{condvstemp}).

The free energy of the system is determined by the action
(\ref{action}) evaluated on-shell $\Omega=-T\,S_{\rm os}$ plus
possible boundary counterterms \cite{skenderislect}. For the
present case the regularized action was presented in
\cite{Herzog:2008he}. We shall proceed along those lines to compute
the free energy of the physical configuration we are interested in.
Substituting the equations of motion (\ref{eom1})-(\ref{eom3}) into
the action (\ref{action}) one finds
\be
S_{0}=\int d^3x\left(
{r^2\over2}\,A_t\,A_t'-{f\over2}\,A_x\,A_x'-r^2\,f\,\psi\,\psi'
\right)\Big|_{r=\infty}+\int d^4x\left( \psi^2\,A_x^2-{r^2\over
f}\,\psi^2\,A_t^2 \right)~,
\label{osaction}
\ee
which is the unregularized on-shell action (the prime means
derivative with respect to $r$). This action consists of three
boundary terms resulting from the kinetic terms of the temporal and
spatial components of the gauge field, and the scalar, respectively;
plus a bulk contribution coming from the interaction terms. From the
asymptotic behavior of the fields (\ref{exp}) it follows that only the
boundary term corresponding to the scalar field $\psi$ is divergent,
and thus we need to add the corresponding counterterm. Moreover, one
must specify the boundary conditions which are  imposed at infinity on
the various fields. In our case, one should add boundary terms which
take us to an ensemble where $\psi^{(2)}, A_t^{(0)}$ and $A_x^{(1)}$
are held fixed, corresponding via eq.~(\ref{physquant}) to $\langle
O^{(2)}\rangle$, the chemical potential $\mu$ and the current
$j_x$. All in all, the boundary term that does the whole job reads
(see \cite{marolf} for a rigorous analysis)\footnote{Varying $A_\mu$
in the bulk yields a boundary term $\sim A_\mu'\delta A_\mu$, which
provides a good variational principle for a boundary condition $\delta
A_\mu|_{r=\infty}=0$. This corresponds to an ensemble where we are
keeping fixed the asymptotic value of $A_\mu$, which in our case means
fixed chemical potential ($\sim A_t|_{r=\infty}$) and fixed source of
the current ($\sim A_x|_{r=\infty}$). One can go to an ensemble where
$\delta S\sim A_\mu\,\delta\,A_\mu'$ by adding a boundary term $\sim
r^2 A_\mu\,A_\mu'|_{r=\infty}$.}
\be
\int d^3x\left(r^3\,\psi^2
+2r^4\,\psi\,\psi'+r^2\,A_x\,A_x'\right)\Big|_{r=\infty}~.
\label{cttrms}
\ee
Substituting the behavior of the fields written in eq.~(\ref{exp})
into the regularized on-shell action given by the sum of the
contributions (\ref{osaction}) and (\ref{cttrms}) yields the following
expression for the free energy
\bea
&&{\Omega(\mu,j_x,\langle O_2\rangle)\over T^3\,V}= \nn \\ &&-\left[
{1\over2}\left( A_t^{(0)}\,A_t^{(1)}+A_x^{(0)}\,A_x^{(1)} \right)
-\psi^{(1)}\,\psi^{(2)} \right] -\int dr\left( \psi^2\,A_x^2-{r^2\over
f}\,\psi^2\,A_t^2 \right)\, ,
\label{freeeng}
\eea where $V$ stands for the volume of the system.

\begin{figure}
\begin{center}
\includegraphics[width=1\textwidth]{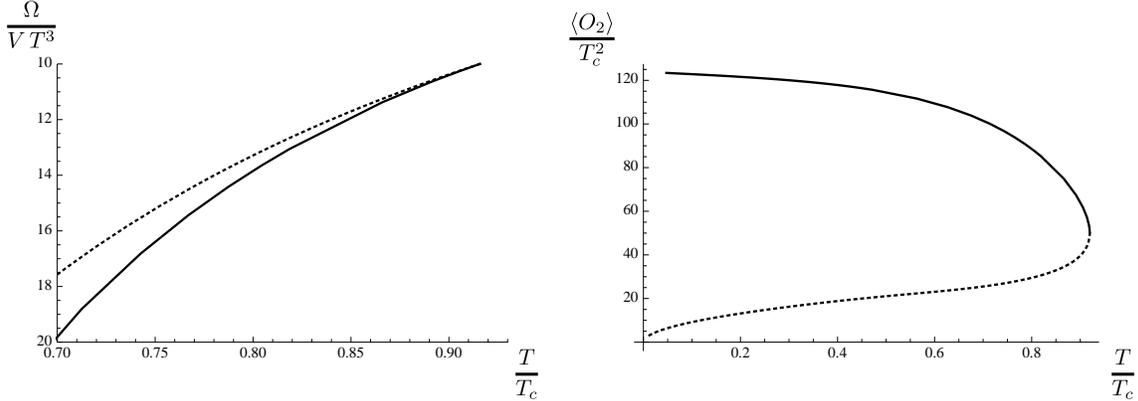}
\end{center}
\caption[Unflavor]{\small For a value of the current $j_x/T_c^2=2.9$,
we plot the free energy of the superconducting phase zooming in on the
region in which $T$ is closer to its maximum value (where the plot
turns back). The dashed line corresponds to the points with a lower
value of the condensate at a given temperature. We show on the right
the corresponding plot of $\langle O^{(2)}\rangle$ versus $T$. One can
see that the lower branch (dotted line) corresponds indeed to states
with larger free energy and thus metastable.}
\label{freengfig}
\end{figure}

We can now compute the free energy of the superconducting states
making up the plot in figure \ref{condvstemp} and confirm that the
lower branch (dotted line) is metastable. This is shown in figure
\ref{freengfig}. Notice, however, that we cannot determine precisely
at which value of the temperature the phase transition occurs. In
order to know this, we would need to compare the free energy of the
superconducting state with the free energy of the normal state at the
same value of the current. Unfortunately, within the minimal framework
we are using, it does not seem possible to describe such a normal
state. Let us elaborate a bit more on this.
Naively, the first thing one could try to do is to look for a solution
with non-trivial $A_t$ and $A_x$ but vanishing scalar ($\psi=0$). However, as noticed in 
\cite{Basu:2008st}, the only such solution satisfying regularity conditions at the horizon  
has $A_x=0$ identically.
%As
%noticed in \cite{Basu:2008st}, such a solution does not exist ($A_x$
%diverges at the horizon and the action becomes infinite there). 
This result should be expected on physical grounds. In the normal state the
dual system is no longer superconducting and thus one expects that in
the absence of an electric field the current must vanish (the DC
conductivity is now finite). One could then try to switch on a
background electric field in the $x$ direction through the addition of
a contribution of the form $-E\cdot t$ to $A_x$. However, in the
Maxwell action having $F_{tx}=-E$ does not modify the equation of
motion for $A_x$ which is then still divergent. One would expect that
using a richer model describing non-linear interactions between the
bulk fields could solve the problem. Indeed, as shown in
\cite{metallic} and more recently in \cite{Hartnoll:2009ns}, a DBI
action does provide a solution with non-vanishing current in the
normal state. There the conductivity was computed and  found to depend
both on the electric field and the charge density. Yet it is not clear
how to implement the scalar condensation corresponding to the
superconducting phase in this scenario \footnote{For a
$p$-wave superconductor such a DBI construction is possible, see for
instance \cite{Ammon:2009fe}.}. A different approach, closer in
spirit to the model we are dealing with here, consists in going beyond
the probe approximation, thus looking for a solution which in the
normal state would correspond to an asymptotically AdS charged black
hole with vector hair (with $A_t(r)$ and $A_x=-E\cdot t+h(r)$).

Let us emphasize that although within the probe approximation regime
we cannot compute the free energy of a normal state with current, the
conclusion about the phase transition being first order is robust. It
is clear from our computations (figure \ref{condvstemp}) that at the
maximum temperature the value of the condensate is different from zero
and then it must jump during the phase transition. If we were able to
compute the free energy of the normal state it may be that the phase
transition  would occur at a value of the temperature somewhat lower
than the maximum value. However, as we see from the plot, the
condensate would still be different from zero at that point. Simply,
the superconducting state would be metastable from the actual
temperature of the phase transition up to the maximum temperature (as
it happens for instance in \cite{Basu:2008st,Herzog:2008he}).

%%%%%%%%%%%%%%%%%%%%%%%%%%%%%%%%%%%%%%%%%%%%%%%%%%%%%%%%%%%%%%%%%%%%%%%%%%%%%%%%%%%%%%%%%%%%%%
\subsection{Current and velocity}
\label{jvsvelsec}

\begin{figure}
\begin{center}
\includegraphics[width=1\textwidth]{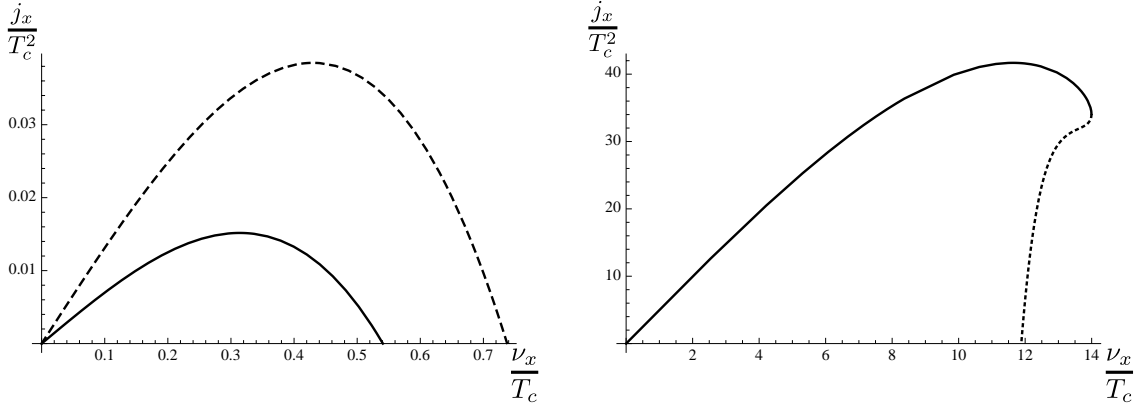}
\end{center}
\caption[Unflavor]{\small Plots of the current $j_x$ versus the
superfluid velocity $\nu_x$ at fixed temperature.  On the left panel
we show the results for two temperatures close to the critical
temperature: $T=0.998\,T_c$ (solid line)  and $T=0.996\,T_c$ (dashed
line). On the right we present the curve we find for $T=0.244\,T_c$,
the dashed line corresponds to  metastable states since they are
solutions with higher free energy than their counterparts with the
same value of the current  or the velocity.}
\label{jvsvx}
\end{figure}

In this section we study the relation between the current $j_x$ and
the superfluid velocity $\nu_x$. As we explained at the beginning of section
\ref{resultsect}, the integration of the equations
(\ref{eom1})-(\ref{eom3}) results in a two-parameter family of
solutions, which we chose to parametrize in terms of $T$ and
$j_x$. This means that once $T$ and $j_x$ are fixed all other physical
quantities of interest are determined up to a discrete choice, in
particular also the superfluid velocity $\nu_x$. We will now fix $T$
and obtain a one-parameter curve of solutions relating $j_x$ and
$\nu_x$. The result is presented in figure \ref{jvsvx} for several
values of $T$. Close to the critical
temperature (left panel) the relation is an upside down paraboloid which becomes
smaller as the temperature approaches $T_c$ (eventually shrinking to a
point for $T=T_c$). On the other hand, for low temperatures (right panel) the
relation between $j_x$ and $\nu_x$ is linear almost all the way up to
a given maximum velocity above which the superconducting state exists
no more.

In general our results match nicely with what is known about the
relation between the current and the superfluid velocity in thin
superconducting films \cite{Tinkham}. As we comment below, though, one
qualitative difference with respect to BCS superconducting films is
that at low temperatures at the maximum superfluid velocity the current is non-zero.

For thin films at temperatures close to $T_c$, where the GL
model is reliable,  the $j_x$ versus $\nu_x$ curve has exactly the
same features as the one in the left panel of figure \ref{jvsvx}. As
we will now explain this is responsible for the different phase
transition pattern one finds when working at fixed current or at fixed
superfluid velocity, respectively. For every value of the current
there are two values of the superfluid velocity. The current is
clearly zero at zero velocity, but also at the maximum value of the
superfluid velocity the current falls to zero because the condensate
vanishes at this velocity (this holds also in the present holographic
model, see \cite{Basu:2008st,Herzog:2008he}).  Therefore, at the
maximum value of the current the superfluid velocity is not at its
maximum and the condensate has a finite non-zero value. This means
that if one increases the current a bit further there is no
corresponding value of the superfluid velocity in the superconducting
phase. Then, the system passes into the normal phase with the
condensate jumping a finite distance to zero, which, as already
explained, is a signal of a first order phase transition.  At the
critical temperature this argument breaks down, as no current is
possible in the superconducting phase and there is no
discontinuity. There the phase transition is second order.

At low temperatures the relation between $j_x$ and $\nu_x$ for
thin films is linear from zero up to a maximum velocity, the depearing
velocity, at which the current falls steeply to zero
\cite{Tinkham}. As we see on the right plot of figure \ref{jvsvx} we
find this linear behavior for a long range of currents. However,
unlike BCS films, we also find that at the maximum value of the
superfluid velocity the current is non-vanishing. According
to the analysis in the previous section, for that value of the current
the condensate is non-zero and hence if one increases the superfluid
velocity the condensate will jump to zero. Hence the phase transition is 
first order. This agrees with the result of \cite{Basu:2008st,Herzog:2008he} 
where it is found that the phase transition at low temperatures is indeed 
first order.

As already noticed, for each value of the current $j_x$ one finds two
solutions with different values of the velocity $\nu_x$ and so two
values of the condensate $\langle O_2\rangle$. As shown in figure
\ref{freengfig}, we have found that the free energy calculated from
the gravity solution using eq. (\ref{freeeng}) is always lower for the
configuration in which the magnitude of $\langle O_2\rangle$ is higher.

%%%%%%%%%%%%%%%%%%%%%%%%%%%%%%%%%%%%%%%%%%%%%%%%%%%%%%%%%%%%%%%%%%
\subsection{Critical current and critical condensate}
\label{checks}

In this section we will discuss two further results of our holographic
analysis. Recall that our model aims to describe a
thin superconducting film and that it is expected to be reliable for
the whole range of temperatures (except for very low temperatures, where
the backreaction needs to be taken into account).  For temperatures near
$T_c$ the GL theory is expected to give an accurate description of
such physical system and therefore we have to compare with GL in
this regime. It is important in the GL derivation that the film is thin, as
this allows one to ignore the free energy contribution of the magnetic
field generated by the current.  As our magnetic field is
non-dynamical, it will not be generated by a current, and so it will
not contribute to the free energy. This makes the comparison between
holographic superconductors, which are inherently ungauged, with GL
model particularly sound for thin films. On the other hand, for
temperatures far below the critical temperature our results give new insights
on the phase diagram of holographic superconductors.

\begin{figure}
\begin{center}
\includegraphics[width=0.95\textwidth]{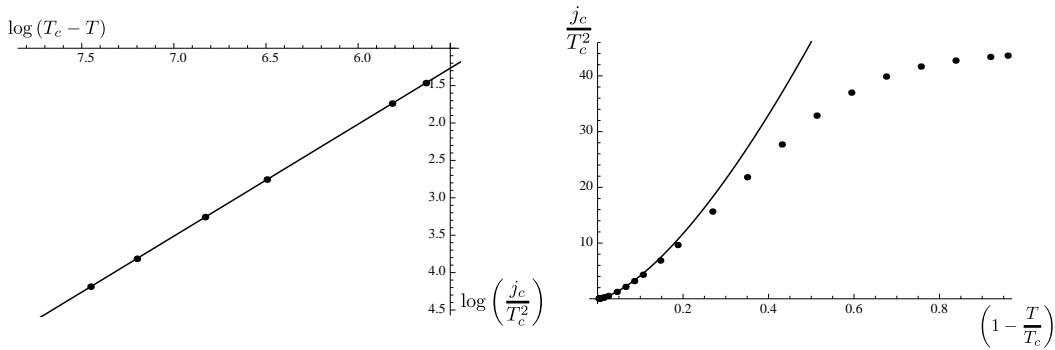}
\end{center}
\caption[Unflavor]{\small
Plot of the critical current versus the temperature. The left panel shows a log-log plot from which we can read-off 
the critical exponent, getting $1.497$, which agrees with the expected GL scaling of 3/2 within our numerical precision. 
The right panel shows the departure from GL scaling (solid line) at low temperatures.}
\label{jcandratfig}
\end{figure}

GL theory predicts that near $T_c$ the critical current $j_c$ is
proportional to $(T_c-T)^{3/2}$. As illustrated in figure
\ref{jcandratfig} we find that this scaling is indeed obeyed by
holographic superconductors for temperatures close to $T_c$. On the other hand, 
at low temperatures our results differ appreciably from GL scaling. This is to be 
expected. For large currents the temperature at which the phase transition occurs  
is appreciably lower than $T_c$, and hence far from the regime where the GL effective description is valid. 
Moreover, being the phase transition first order, one does not expect any power-law scaling. 
Notice that we are defining the
critical current to be the highest current at which the
superconducting solution exists, at a given temperature. As we discussed in
section \ref{freeensect}, it might be that the phase transition
happens for a lower value of the current and thus the value we are
considering would correspond to a metastable state. Nevertheless, at
temperatures close to $T_c$ one can reasonably expect than the
corrections to the free energy coming from the very small value of the
current are almost negligible and the critical current agrees with the
maximum allowed value. The fact that under this assumption the result
got for the holographic superconductors agrees with GL is an a
posteriori reassurance. Being more conservative, one should take our
result as an upper bound, especially for the large current
(corresponding to low temperature) regime: in other words, at a given
temperature, the corresponding critical current would be at most
equal to the one predicted by the plot in figure \ref{jcandratfig}.

\begin{figure}
\begin{center}
\includegraphics[width=0.65\textwidth]{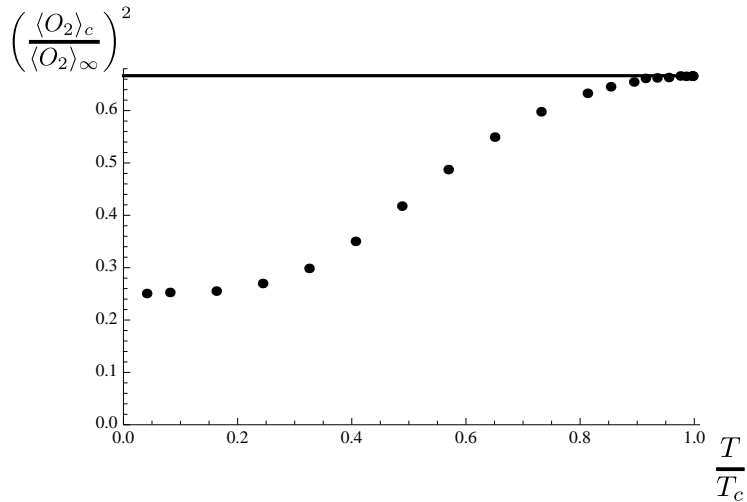}
\end{center}
\caption[Unflavor]{\small
Plot of the ratio $(\langle O_2\rangle_c\, / \,\langle O_2\rangle_\infty)^2$ versus the temperature. The solid
line corresponds to the value of $2/3$ predicted by the GL theory.}
\label{jcandratfig2}
\end{figure}

A second prediction of the GL theory is that, at any fixed temperature, the norm of the condensate monotonically
decreases with respect to the velocity from its maximum value $\langle O_2\rangle_\infty$. The critical current
is reached before the maximum velocity, when the norm of the condensate has an intermediate value
$\langle O_2\rangle_c$. More precisely one has
\be
\left(\frac{\langle O_2\rangle_c}{\langle O_2\rangle_\infty}\right)^2=\frac{2}{3}~.
\label{dueterzi}
\ee
We have found numerically that this relation is also satisfied for holographic superconductors at temperatures near
the critical temperature $T_c$. This can be seen in figure \ref{jcandratfig2} where we plot the ratio (\ref{dueterzi})
versus the temperature. Again, away from $T_c$ the behavior changes sensibly. Notice that the
same warning about our inability to determine exactly the critical current applies here and therefore,
especially in the region $T\ll T_c$, the points in figure \ref{jcandratfig2} should be considered as a lower bound
for the ratio $(\langle O_2\rangle_c\, / \,\langle O_2\rangle_\infty)^2$. Notice that the ratio
goes to a constant at zero temperature.

%%%%%%%%%%%%%%%%%%%%%%%%%%%%%%%%%%%%%%%%%%%%%%%%%%%%%%%%%%%%%%%%%%%%%%%%%%%
\subsection{Conductivity}

In this section we will study the AC conductivity of the system and characterize its dependence on the DC current.
To compute the conductivity one must consider an electromagnetic perturbation on top of the hairy black hole solution.
This is easy for a perturbation along the direction
orthogonal to the current (\ie \ a perturbation of $A_y$), since it decouples from other perturbations of the gauge
vector or the scalar field. Conversely, a perturbation of $A_x$ couples to perturbations of $A_r$ and $\psi$ and
hence the computation of the conductivity along the direction parallel to the current becomes more involved
and we will not attempt to do that here.

The equation of motion for a zero-momentum perturbation $\delta A_y=e^{-i\,\omega\,t}A_y(r)$ takes the form
\be
\partial_r\left(f\,\partial_r A_y \right) + \left({\tilde\omega^2\over f} - 2\psi^2\right)A_y=0~,
\label{ayperturb}
\ee
where we have applied again the rescalings given in eq. (\ref{newfields}) and defined
$\tilde\omega=3/(4\pi)\,\omega/T$. Notice that this is the same equation as considered in \cite{Hartnoll:2008vx},
but the background solution for the scalar $\psi$ is different now, in particular it depends on the current.
The boundary asymptotics ($r\to\infty$) of $A_y$ take the form
\be
A_y=A_y^{(0)}-{A_y^{(1)}\over r}+ O(r^{-2})~.
\label{ayasympt}
\ee
The conductivity is given by the zero-momentum retarded current-current correlator which by using the AdS/CFT dictionary
can be calculated in terms of solutions satisfying ingoing wave boundary conditions at the horizon
\cite{Son:2002sd}. In fact we recover Ohm's law on the boundary
\be
\sigma_y(\omega)={\langle J_y\rangle\over E_y}=i{A_y^{(1)}\over\omega\,A_y^{(0)}}~,
\label{cndctvty}
\ee
where we have taken into account that $A_y^{(0)}$ is introducing a background potential on the boundary and thus an
electric field $E_y=-\partial_t\,A_y^{(0)}$.

\begin{figure}
%[ht]
%\begin{center}
{\hskip -1cm\includegraphics[width=1.05\textwidth]{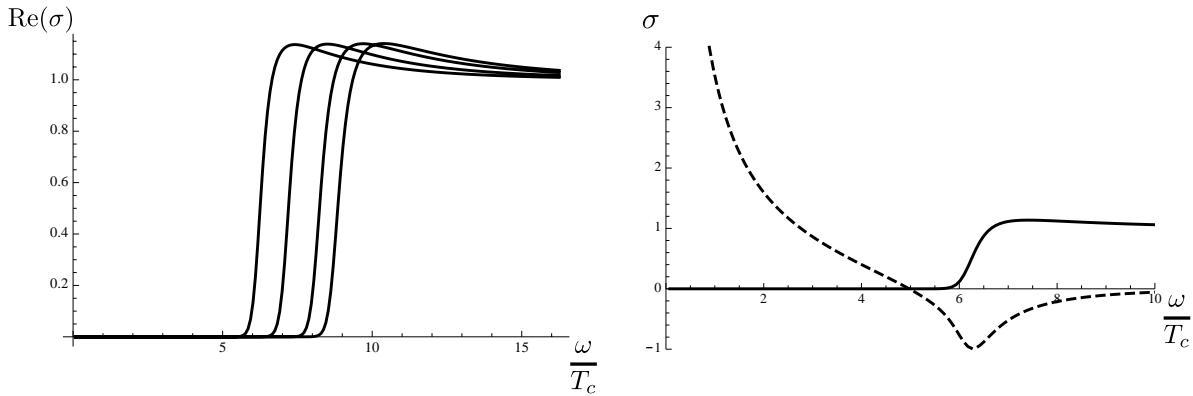}}
%\end{center}
\caption[Unflavor]{\small
On the left we plot the real part of the conductivity versus the frequency for several values of the current at $T=0.04\,T_c$.
From left to right: $j_x/T_c^2=43.6\,,\;41.3\,,\;29.6\,,\;2\cdot 10^{-6}$. The leftmost curve corresponds to the maximum current
at this temperature. On the right we show both the real (solid line) and imaginary (dashed) part of the conductivity as a
function of the frequency again at $T=0.04\,T_c$ and with $j_x/T_c^2=43.6$, the largest allowed current at that temperature.}
\label{accond}
\end{figure}

By solving numerically eq. (\ref{ayperturb}) with infalling boundary conditions at the horizon ($r=1$) we can compute the
conductivity as a function of the frequency $\omega$ at given values of temperature and current.
In figure \ref{accond} we show the results obtained at a low temperature ($T=0.04\cdot T_c$) for different values of the current.
The AC conductivity displays the features already observed in \cite{Hartnoll:2008vx}. At large frequencies it approaches
a constant, a characteristic of theories with $AdS_4$ duals \cite{herzads4}.
On the other hand, at $\omega=0$ we expect a delta function in ${\rm Re}(\sigma)$,
a fact confirmed, via the Kramers-Kronig relations, through the presence of a pole in the imaginary part of $\sigma$ at $\omega=0$.
Finally, we see that for small enough frequencies, within our numerical precision, ${\rm Re}(\sigma)$ vanishes. This
gap can be parametrized in terms of a critical frequency $\omega_g$.
As we show in figure \ref{accond} there is a minimum of ${\rm Im}(\sigma)$ around the point where ${\rm Re}(\sigma)$ becomes
non-zero. Then, following \cite{Horowitz:2008bn}, we define $\omega_g$ as the frequency minimizing the imaginary part of the conductivity.

\begin{figure}
%[ht]
\begin{center}
\includegraphics[width=0.5\textwidth]{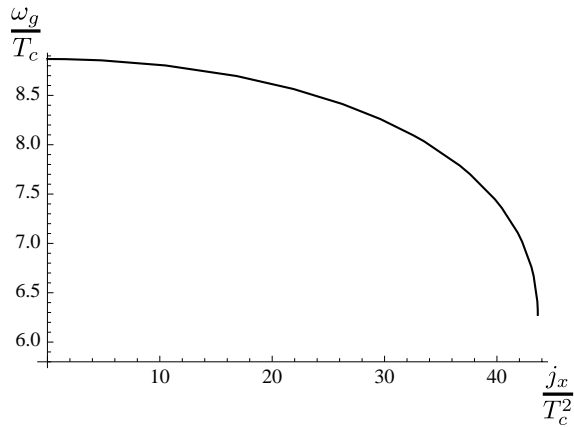}
\end{center}
\caption[Unflavor]{\small
Plot of $\omega_g$ as a function of the current. At zero current we find $\omega_g/T_c=8.87$, while at
the maximum current $\omega_g/T_c=6.28$.}
\label{omgap}
\end{figure}

For weakly coupled superconductors the gap is predicted to be $\omega_g/T_c = 3.5$ at $T=0$ \cite{Tinkham}. In the original model
of \cite{Hartnoll:2008vx} the gap was found to be $\omega_g/T_c \approx 8$ and it seems that such a high value holds quite generically
for holographic superconductors. This has been seen as an indication that holographic superconductors are indeed strongly coupled.
In figure \ref{omgap} we plot $\omega_g$ as a function of the current. In the region of very low current we recover the result
$\omega_g/T_c \approx 8$. As we increase the current, $\omega_g$ decreases continuously until we reach the maximum current where
it has a finite value. This is consistent with the phase transition being first order at that point. The condensate is non-vanishing
and thus we expect $\omega_g$ to be also different from zero.

In weakly coupled superconductors a definite relation exists between $\omega_g$ and the energy gap $\Delta$ at zero temperature,
$\omega_g = 2 \Delta$, $\Delta(T)$ being the minimum energy required for charged excitations at a given temperature $T$. In
strongly coupled superconductors one does not expect the gap to necessarily satisfy this relation so one could have wondered what
the relation is for holographic superconductors. However, as noticed in \cite{Hartnoll:2008kx,Peeters:2009sr,Horowitz:2009ij} and
recently reviewed in \cite{Horowitz:2010gk}, holographic superconductors are not hard-gapped, in general, and a non-zero conductivity is present
even at small frequencies, though exponentially suppressed (for recent work on hard-gapped holographic superconductors see for instance
\cite{Basu:2009vv}, and \cite{Aprile:2010yb} for a recent
discussion on this point). Indeed, computations of the temperature dependence of the specific heat \cite{Peeters:2009sr} showed that
holographic superconductors behave similarly to some strongly coupled superconductors as heavy fermion compounds: the specific heat
does not vanish exponentially at low temperature (this being a consequence of, and hence an indication for, the existence of an
energy gap), but as a power law.

%%%%%%%%%%%%%%%%%%%%%%%%%%%%%%%%%%%%%%%%%%%%%%%%%%%%%%%%%%%%%%%%%%%%%%%%%%%%%%%%%%%%%%%%
%%%%%%%%%%%%%%%%%%%%%%%%%%%%%%%%%%%%%%%%%%%%%%%%%%%%%%%%%%%%%%%%%%%%%%%%%%%%%%%%%%%%%%%%%%%%%%%%%%
\section{Conclusions and outlook}

In this paper we have considered a simple holographic model of a thin
superconducting film with DC current. We focused on the modifications
that the presence of a DC current induces on the thermodynamics as
compared to the same holographic model with vanishing current,
originally studied in \cite{Hartnoll:2008vx}. Most notably, the phase
transition becomes first order for any finite value of the
current. Moreover, the conductivity gap becomes a function of the
current, too: the frequency gap diminishes as one increases the
current but never reaches zero before the phase transition occurs,
in agreement with the phase transition being first
order. Other results we obtain nicely agree with expectations for thin
superconducting films, such as the relation between the current and
the superfluid velocity, both at low and high (that is near to $T_c$)
temperatures. The only qualitative difference is that at
sufficiently low temperatures and high superfluid velocities, the
velocity no longer uniquely determines the current.

Ideally one would like to go beyond the probe approximation. Besides
leading to better control over the very low temperature regime, this is
necessary in order to obtain a holographic description of the normal phase.
To describe the superconductor in the normal phase with a
DC current one should switch on an external electric field, which is
needed to keep a constant current in the normal phase, since there we
have a non-vanishing resistivity. However, as we already noticed, only
the full system of coupled scalar Maxwell-Einstein gravity equations
could in principle allow for a non-trivial but meaningful (that is
non-singular) solution. This implies that one should go for a fully
back-reacted analysis.

%%%%%%%%%%%%%%%%%%%%%%%%%%%%%%%%%%%%%%%%%%%%%%%%
\subsection{Towards holographic Josephson junctions}

Our primary motivation for this work was the observation that HTS's
have typically a layered structure and may, according to the
highly successful Lawrence-Doniach model \cite{lawrence}, be
approximated by superconducting films separated by Josephson
junctions. While the present model can be seen as a first
step in this direction, one would like to find a complete holographic
description of a Josephson junction. Let us elaborate a bit on this.

One kind of Josephson junction that appears particularly amenable to
a holographic construction is the S-c-S
junction. Such a junction is composed entirely of the same
superconducting material, but the superconductor is thinner at the
junction, and so for example will have a lower critical current.
Imposing a space-dependent metric in the boundary theory, one
might be able to cook-up a dual model with varying critical
current. This might seem quite ad hoc in a bottom-up context but
it might possibly arise quite naturally in a string theory
context. For example, one may consider M-theory compactified on a 7-dimensional Sasaki-Einstein
manifold fibered over $AdS_4$. The critical current, like all quantities in a holographic superconductor,
depends on the compactification geometry. Then, to construct a junction, one
may merely need to vary the moduli of the 7-manifold in the region
corresponding to the junction, that is, over a finite interval in one
of the field theory directions.  An array of junctions would then
correspond to moduli that vary periodically in one field theory
direction.

Ideally such a spatial dependence of the moduli will be a solution of
the supergravity equations of motion.  One hope of realizing such a
solution is as follows.  Each Josephson function may correspond to a
brane extending from the boundary to the horizon and along the field
theory directions parallel to the Josephson junction, and also
potentially wrapping some cycle of the compactification manifold.  One
may then hope that by correctly choosing this cycle, one may engineer
a geometry in which the backreaction of the brane causes the desired
deformation of the compactification manifold, reducing the critical
current and therefore forming an S-c-S junction. A Lawrence-Doniach
HTS may then correspond to an array of such branes.

%%%%%%%%%%%%%%%%%%%%%%%%%%%%%%%%%%%%%%%%%%%%%%%%%%%%%%%%%%%%%%%%%%%%%%%%%%%%%%%%%%%%%%%%%5
\section*{Acknowledgments}

We thank Irene Amado, Francesco Aprile, Pallab Basu, Stefano Cremonesi, Michele Fabrizio, Chris Herzog, Chethan Krishnan, Gary Horowitz,
Alessandro Laio, Sergej Moroz, Andy O'Bannon, Kostas Skenderis, Dam Son, Andrea Trombettoni and Andrea Wulzer for useful
discussions and/or email correspondence at various stage of this work.

%%%%%%%%%%%%%%%%%%%%%%%%%%%%%%%

\end{document}